\documentstyle[draft]{npbproc}

\def\cgt{CGT\ }
\def\cym{CYM\ }

\def\null{\hbox{}}

\def\linhl{\hrule height .2mm width 10.2mm depth 0mm}
\def\linhc{\hrule height .2mm width 10mm depth 0mm}
\def\linv{\vrule height 9.8mm width .2mm depth 0mm}
\def\co{\vbox{\linhl\hbox to 10.2mm{\linv\hfill\linv}\linhl}}
\def\cc{\vbox{\linhc\hbox to 10mm{\linv\hfill}\linhc}}
\def\cu{\vbox{\hbox to 10.2mm{\linv\hfill\linv}\linhl}}
\def\cl{\vbox{\hbox to 10mm{\linv\hfill}\linhc}}
\def\rin{\hbox{\cl\cl\cl\cl\cu}}
\def\rinb{\hbox{\cc\cc\cc\cc\co}}
\def\cle{\vbox{\hbox to 10mm{\linv\vbox to 9.8mm{
\vskip .3mm\hbox to 9.8mm{\hfill $\epsilon $}\vfill}}\linhc}}
\def\clm{\vbox{\hbox to 10mm{\linv\vbox to 9.8mm{
\hbox to 9.8mm{\hfill $-\epsilon $}\vfill}}\linhc}}
\def\ceh{\hbox to 10mm{\hfill$\epsilon$}}
\def\gcehp{\hbox{\ceh\ceh\ceh\ceh}}
\def\rinp{\vtop{\null\vskip .3mm\gcehp}}
\def\cmh{\hbox to 10mm{\hfill$-\epsilon$}}
\def\gcehs{\hbox{\cmh\ceh\cmh\ceh}}
\def\rinVI{\vtop{\null\vskip .3mm\gcehs}}
\def\rinIV{\hbox{\cle\cle\cle\cl\cu}}
\def\rinV{\hbox{\clm\clm\clm\cl\cu}}
\def\rinVIII{\hbox{\clm\cle\clm\cl\cu}}
\def\rinIX{\hbox{\cle\clm\cle\cl\cu}}
\def\cthp{\vbox{\hbox{\hskip 10mm $\theta = (0,\pi)$}\vskip .3mm}}
\def\cths{\vbox{\hbox{\hskip 10mm $\theta = (\pi,\pi)$}\vskip .3mm}}
\def\cev{\vbox to 10mm{\vskip .3mm\hbox to 6mm{\hfill $\epsilon$}\vfill}}
\def\cnev{\vtop{\null\vbox to 5mm{\vskip .3mm
\hbox to 6mm{\hfill $\epsilon$}\vfill}}}
\def\cmv{\vbox to 10mm{\hbox to 6mm{\hfill $-\epsilon$}\vfill}}
\def\riv{\vbox{\cev\cmv\cev\cmv\cnev}}
\def\flev{$\left\uparrow\vcenter to 20mm{}\right.$}
\def\cflev{\vbox{\hbox{\flev}\null}}
\def\cfleh{\hbox to 20mm{\rightarrowfill}}
\def\figpa{\hbox{\cflev\riv\vbox{
\cthp\rinb\rin\rin\rinIV\rinV\vtop{\null\rinp\cfleh}}}}
\def\figpb{\hbox{\cflev\riv\vbox{
\cths\rinb\rin\rin\rinVIII\rinIX\vtop{\null\rinVI\cfleh}}}}
\def\fig{\offinterlineskip\large \hbox to \hsize{\figpa\hfill\figpb}}

\begin{document}
\title{MASSLESS DECOUPLED DOUBLERS: CHIRAL YUKAWA MODELS AND
CHIRAL GAUGE THEORIES%
\thanks{Talk presented at the topical workshop \lq\lq Non perturbative aspects
of Chiral Gauge Theories\rq\rq, Accademia Nazionale dei Lincei, Roma,
9-11 March,1992.}}

\author{J.L. Alonso, J.L. Cort\'es, F. Lesmes}%
\address{Departamento de F\'{\i}sica Te\'orica. Universidad de Zaragoza \\
         50009 Zaragoza, Spain}
\author{Ph. Boucaud}%
\address{Laboratoire de Physique Th\'eorique et Hautes Energies. \\
         Universit\'e Paris XI, 91405 Orsay Cedex, France}
\author{E. Rivas}%
\address{Department of Physics, Washington University, \\
         St. Louis, MO 63130, USA}

\runtitle{Massless decoupled doublers \ldots}
\runauthor{J.L. Alonso et al.}

\begin{abstract}
We present a new method for regularizing chiral theories on the lattice. The
arbitrariness in the regularization is used in order to decouple massless
replica fermions. A continuum limit with only one fermion is obtained in
perturbation theory and a Golterman--Petcher like symmetry related to the
decoupling of the replicas in the non--perturbative regime is identified.
In the case of Chiral Gauge Theories gauge invariance is broken at the level
of the regularization, so our approach shares many of the characteristics of
the Rome approach.
\end{abstract}

\maketitle

\section{INTRODUCTION}
In order to make some progress on a rigorous definition of quantum field
theories and study their non--perturbative effects a crucial step is to use a
lattice as regulator. In the case of a $\rm \underline C$hiral
$\rm \underline G$auge $\rm \underline T$heory, any attempt in this
direction [1,2] is based on a solution of the well known
difficulties (doubling problem) one finds for theories with fermion fields on
the lattice. Several ways of dealing with this problem, which have been
proposed recently [3], have been reviewed in references [1] and [4], so that
our interest here is concentrated mainly on our proposal for the decoupling of
the doublers.

\subsection{Philosophy of our approach}
We will start by saying the philosophy of our approach.

As replica fermions look inevitable, let us try to live with decoupled but
massless, (i.e. harmless), replica fermions. This has to be contrasted with the
usual ways of decoupling the replica by giving them a mass of the order of the
cut--off. Our first tentatives were the works
in reference [5] and some results in two spacetime dimensions can be found in
reference [6].

To implement this idea, we use the arbitrariness in the regularization in order
to couple, by hand,
{\it in the action,}
only one of the $2^d$ fermions to the
scalars, in the case of $\rm \underline C$hiral $\rm \underline Y$ukawa
$\rm \underline M$odels, and to the gauge fields and the
scalars, in the case of CGT. In the case of CGT this \lq\lq decoupling by
hand\rq\rq
can be done at the expense of working with a non gauge invariant
regulatization,
so our approach shares many of the characteristics of the Rome approach
[7].

Given that dimensional regularization, the pa\-ra\-digm of a gauge invariant
preserving regularization, does not work very well in the case of CGT
[8], it seems natural to investigate the chances offered by a
gauge non--invariant regularization on the lattice.

\subsection{Properties}
In both, CGT and CYM, we are able to describe fermions with arbitrary chiral
content (a chiral field, $\Psi_L$ for instance, does not need the presence of
a complementary one, $\Psi_R$). Also our procedure involves a minimal field
content (no mirror fermions, for instance).

For the case of CYM the regularization method keeps all symmetries of the
continuum: hermiticity, invariance under discrete rotations and translations
and (global) chiral symmetry (see section 3 for some comments about the
symmetries of the quantum theory).

Besides, no extra tuning is needed in order to give finite mass to the physical
fermions (and to decouple the replicas). Also the global chiral symmetry we
have
{\it (in the case of a CGT we lose only the corresponding local symmetry)}
avoids
the ocurrence of a mass counterterm for the fermion field: the masses of the
fermions are protected.

\subsection{State of the art}
 a) We have proved perturbatively that in the continuum limit only one
fermion is coupled to the scalars, in the case of a CYM, and to the scalars
 and the gauge fields, in the case of a CGT (in this case at one loop level).

b) We have (partially) found the phase diagram of a CYM.

c) We have a nice Golterman--Petcher like symmetry [9]
related to the decoupling of the replicas in the non perturbative regime of
both
CGT and CYM.

\section{THE LATTICE ACTION FOR A CGT AND THE DECOUPLING SYMMETRY}
Concerning a CGT we have done little but proposing our regularization. The
action has five terms (for details of notation, see J.~Smit at the Capri
Conference [1]),
\begin{eqnarray}
I(\Psi,\Phi,U)&=&I_U + I_B(\Phi,U) + I_F(\Psi) \nonumber \\
              & &\mbox{} + I_{INT}(\Psi,U)+ I_Y(\Phi,\Psi),
\end{eqnarray}
where,
\begin{eqnarray}
I_U &=& {1 \over g^2a^{4-d}} \sum_{x,\mu,\nu} Tr[U_{L x \mu \nu} +
U_{L x \mu \nu}^+] \nonumber \\
    & &\mbox{} + ( L \leftrightarrow R ),
\end{eqnarray}
is the usual plaquette term for the gauge fields made out of link variables
$U_{x \mu}$,
\begin{eqnarray}
I_B(\Phi,U) &=& - \sum_x {1 \over 2} Tr [ \Phi_x^+\Phi_x +
\lambda (\Phi_x^+\Phi_x - 1)^2] \nonumber \\
            & &\mbox{} + {k \over 2} \sum_{x,\mu} Tr[\Phi_{x+\hat\mu}^+
U_{L x \mu}^+\Phi_xU_{R x \mu} \nonumber \\
            & &\mbox{} + h.c. \quad],
\end{eqnarray}
is the lattice action for a complex scalar field.
$I_F(\Psi)$, the free action, is,
$$ I_F(\Psi) = I_N(\Psi,U=1), $$
$I_N$ being the naive action,
\begin{eqnarray}
I_N(\Psi,U) &=& {} - a^{d-1}{1\over 2}
\sum_{x,\mu}[\bar\Psi_{Lx}\gamma_\mu U_{Lx\mu}\Psi_{Lx+\hat\mu} \nonumber \\
            & & {} -\bar\Psi_{Lx+\hat\mu}\gamma_\mu U_{Lx\mu}^+\Psi_{Lx}]
\nonumber \\
            & & {} + ( L \leftrightarrow R ).
\end{eqnarray}

The interaction term,
\begin{eqnarray}
I_{INT}(\Psi,U) = I_N(\Psi^{(1)},U) - I_N(\Psi^{(1)},1),
\end{eqnarray}
contains the coupling of the fermions to the gauge fields. (Note that if we put
$U = 1$, then $I_{INT}(\Psi,U) = 0$).

Finally, the Yukawa term is
\begin{eqnarray}
I_Y(\Phi,\Psi) &=& - y \sum_x(\bar\Psi^{(1)}_{Lx}\Phi_x\Psi^{(1)}_{Rx}
\nonumber \\
               & & {} + \bar\Psi^{(1)}_{Rx}\Phi_x^+\Psi^{(1)}_{Lx}).
\end{eqnarray}

{\it The way to implement the decoupling of the replica fermions is based
on the use, in the interaction terms, of the component $\Psi^{(1)}$.}
In momentum space $\Psi^{(1)}$ is given by
\begin{eqnarray}
\Psi^{(1)}(\theta) = F(\theta)\Psi(\theta),\qquad
&     F&(\theta) = \prod_\mu f(\theta_\mu), \nonumber \\
f(\theta) = \cos({\theta\over 2}),  \hfill
&\theta& \in (-\pi,\pi].
\end{eqnarray}

Note that $f(0) = 1 ,\quad f(\pi) = 0$. As we see,
$\Psi^{(1)}$ is the original $\Psi$ field but modulated by a form factor which
is responsible of the decoupling of the replica fermions at the tree level.
This is achieved by imposing that the form factor vanishes for the momenta
corresponding to the replica. We have chosen the most local solution.

In coordinate space the field $\Psi^{(1)}$ corresponds to an average over the
$\Psi$ components defined in an elementary hypercube in the positive
directions,
\begin{eqnarray}
\Psi^{(1)}_x &=:& \int_\theta \exp\lbrace i\theta(x+{1\over2}\sum_\mu\hat\mu)
\rbrace F(\theta)\Psi(\theta)  \nonumber \\
             &= & {1\over2^d} [\Psi_x + \sum_{n=1}^d\sum_{\mu_{i_1}<\dots
<\mu_{i_n}}\Psi_{x+\hat\mu_{i_1}+\dots+\hat\mu_{i_n}}] \nonumber \\
             &= & {1\over2^d} [\Psi_x                 + \Psi_{x+\hat 1}
+ \Psi_{x+\hat 2}+ \Psi_{x+\hat 3} + \Psi_{x+\hat 4} \nonumber \\
             &  & {} + \Psi_{x+\hat 1+\hat 2}
+ \Psi_{x+\hat 1+\hat 3} + \dots ].
\end{eqnarray}

All terms in (1) which include the fermion field break the gauge invariance,
\begin{eqnarray}
\Psi_x     &\longrightarrow& (\Omega_{Lx}P_L + \Omega_{Rx}P_R)\Psi_x,\\
\bar\Psi_x &\longrightarrow& \bar\Psi_x(\Omega_{Lx}^+P_R + \Omega_{Rx}^+P_L),\\
U_{Lx\mu}  &\longrightarrow& \Omega_{Lx}U_{Lx\mu}\Omega^+_{Lx+\hat\mu},\\
U_{Rx\mu}  &\longrightarrow& \Omega_{Rx}U_{Rx\mu}\Omega^+_{Rx+\hat\mu},\\
\Phi_x     &\longrightarrow& \Omega_{Lx}\Phi_x\Omega^+_{Rx}.
\end{eqnarray}

$I_{INT}(\Psi,U)$ breaks gauge invariance because in
$\bar\Psi_{Lx}^{(1)}\gamma_\mu U_{Lx\mu}\Psi_{Lx+\hat\mu}^{(1)}$ the same
gauge variable, $U_{Lx\mu}$, couples $\Psi$'s in different points.
The same argument $(U\leftrightarrow\Phi)$ applies to $I_Y(\Phi,\Psi)$.

As emphasized by the Rome group [7], the model has to be defined
in the presence of the gauge fixing and Fadeev--Popov terms, so that one must
keep such terms also at the non-perturbative level. At the same time one
has to include all the counterterms which correspond to gauge non invariant
terms of dimension less or equal to four required in order to recover a gauge
invariant theory in the continuum limit.

\begin{figure*}
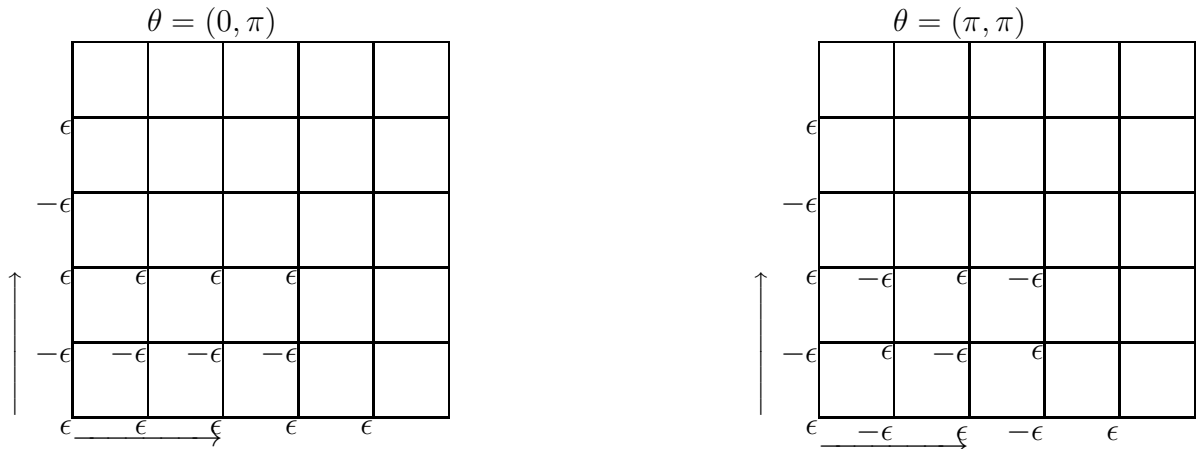

\fig
\caption{Ilustration of transformations (14)}
\end{figure*}

The consistency of the model has been established at one loop level when the
fermion field content is anomaly free [7].

A non-perturbative tuning associated to the quadratically divergent mass term
for {\it each} gauge field is present due to the non--invariance
of the regularization (see L.Maiani, in these proceedings, for comparison with
the Rome approach).
A difference of our method versus that of Rome is that our fermion sector has a
global chiral symmetry which prevents the occurrence of linearly divergent mass
counterterms, of the type $\bar\Psi\Psi$, for the fermion fields. This reduces
the number of non--perturbative tunings needed.

At the same level at which the consistency has been proved (i.e. at one loop),
we have enough $F(\theta)$ factors to assure the decoupling of the replica
fermions. As we will see in the next section, the best proof of this is based
on the Reisz power counting theorem [10] concerning Feynman integrals.

In that section we will state this theorem and it will be used to prove the
decoupling of the replica fermions in CYM, a case in which the problems related
to the gauge non invariance of our regularization (which prevent us to go
beyond one loop) are not present.

Now, we will present an argument of plausibility to argue that the decoupling
works also in the non--perturvative regime. This argument is based on a
symmetry of our action which is a consequence of the use of the component
$\Psi^{(1)}$ in
the interaction term. Actually, the action is invariant under
the $2^d-1$ transformations of the fermion field,
\begin{eqnarray}
\Psi_x     &\rightarrow& \Psi_x + \epsilon_x^{(i)}        ; \qquad
\epsilon_{x+\hat\mu}^{(i)}     = e^{i\theta_\mu^{(i)}}\epsilon_x^{(i)}
\nonumber \\
\bar\Psi_x &\rightarrow& \bar\Psi_x + \bar\epsilon_x^{(i)}; \qquad
\bar\epsilon_{x+\hat\mu}^{(i)} = e^{i\theta_\mu^{(i)}}\bar\epsilon_x^{(i)},
\end{eqnarray}
where $\theta_\mu^{(i)} = 0,\pi$ and at least one component is different from
zero (momenta corresponding to the replica fermions).

An illustration, in two dimensions, of these $2^d-1$ transformations is in
figure 1,
where the value of the translation in the fermionic field in each
point of the lattice is shown for $\theta = (0,\pi)$ and $\theta = (\pi,\pi)$.

The invariance of the interaction terms $I_Y$ and $I_{INT}(\Psi,U)$ follows
from
the invariance of the $\Psi^{(1)}$ component, entering on these terms, under
transformations (14). This invariance results from the fact that the sum of the
$\epsilon$'s in an elementary hypercube is zero. By inspection it is
straightforward to check that the naive term, $I_F(\Psi)$ in (1), is also
invariant under these transformations. In fact it is also invariant under the
transformation $\epsilon_{x+\hat\mu}=\epsilon_x$.

It is normal to have $2^d-1$ transformations of symmetry which are translations
in the fermionic variable if one expects to have $2^d-1$ free massless fermions
[9].

As we have said, we expect that, due to this symmetry,
{\it the decoupling will work also in the non--perturbative regime}

In fact, in the case of a CYM, it is easy to derive the Ward identities
associated to these symmetries [11]. The result,
{\it for finite $a$,} is:

1) The position of the replica fermions in momentum space is not changed by
quantum corrections.

2) The one-particle irreducible fermion Green functions do not see the
doublers,
in the sense that any 1PI amputated Green function vanishes if at least one
external fermion line has $\theta = \theta^{(i)}$.

Consequently, the identification of this symmetry can be used to argue that
also
in a non--perturbative regime the replica fermions decouple,
$\underline{\rm IF}$ a sensible continuum limit exists with only the terms of
the action of the CYM\footnote {Of dimensions less than or equal to four}
(which are those exhibiting the symmetry). Without this
$\underline{\rm IF}$ the proof of the decoupling
{\it in the non--perturbative regime}
of a CYM would be complete.
The $\underline{\rm IF}$ converts the proof in only \lq\lq an argument of
plausibility\rq\rq.

Because this transformation of symmetry acts only on the fermion field
the former conclusion about the decoupling is extensible to a CGT but now the
$\underline{\rm IF}$ is a stronger assumption. In fact, in the case of
a CGT the existence of a sensible continuum limit with only terms of dimension
less or equal to four is an open problem, even perturbatively (as only at
one--loop level has consistency been proved [7]).
To progress on the question of the existence of this limit is one of the
points of this Workshop.

This \lq\lq decoupling symmetry\rq\rq\ is in correspondence with the
Golterman--Petcher
symmetry [9] in the case of the Wilson--Yukawa formulation
of the Standard Model which allows one to prove the decoupling of the
right--handed neutrino in the continuum limit.

\section{CHIRAL YUKAWA MODELS ON THE LATTICE AND DECOUPLING}

The simplest fermion--scalar sector of a \cgt has the following global
$ U(1) \times U(1) $ invariant action,
\begin{eqnarray}
I(\Psi,\Phi) = I_B(\Phi) + I_F(\Psi) + I_Y(\Phi,\Psi),
\end{eqnarray}
where,
\begin{eqnarray}
I_B(\Phi) = {} &-& \sum_x\Phi_x^+\Phi_x \nonumber \\
            {} &+& {k \over 2}\sum_{x,\mu}(\Phi_{x+\hat\mu}^+\Phi_x
+ \Phi_x^+\Phi_{x+\hat\mu})\nonumber \\
            {} &-& \lambda \sum_x(\Phi_x^+\Phi_x - 1)^2,
\end{eqnarray}
\begin{eqnarray}
I_F(\Psi) = {} &-& a^{d-1}{1\over 2}
\sum_{x,\mu}(\bar\Psi_x\gamma_\mu \Psi_{x+\hat\mu} \nonumber \\
            {} &-& \bar\Psi_{x+\hat\mu}\gamma_\mu\Psi_{x}),
\end{eqnarray}
\begin{eqnarray}
I_Y(\Phi,\Psi) &=& - y \sum_x(\bar\Psi^{(1)}_{Lx}\Phi_x\Psi^{(1)}_{Rx}
\nonumber \\
               & & {} + \bar\Psi^{(1)}_{Rx}\Phi_x^+\Psi^{(1)}_{Lx}).
\end{eqnarray}

Now, we are going to prove, in a weak coupling perturbative analysis, using
power counting arguments, that the doublers are decoupled. We would like to
emphasize that this will be a rigorous proof (in contrast with the previous,
{\it non--perturbative,} argument of plausibility).

The proof is based on the Reisz power counting theorem which, essentially,
says \footnote
{The basic idea of the theorem is to bound from below the
denominators of the propagators of a lattice Feynman integral. This, as we
will see, is not possible for the denominator of the naive propagator
(because of the doubling problem).}:

\medskip {\noindent \bf Theorem}

Let $I_F$ be a lattice Feynman integral,
\begin{eqnarray}
I_F = \int d^4k^1 \dots d^4k^L {V(k,q;a) \over C(k;a)},
\end{eqnarray}
where $k^1,\dots,k^L$ are the loop momenta, $q$'s are the external momenta,
$V$ includes all vertex factors and the numerators of the propagators and
$$
C(k;a) = \prod_i C_i(k^i;a).
$$
$C_i(k^i;a)$ are the denominators of the propagators. $C_i$ is required to have
the following properties ( these assumptions, less strong than those originally
assumed by Reisz, were established by L\"uscher [12]):

1. There is a smooth function $G_i$ ($2\pi$--periodic in the momentum
$ak^i$) such that
\begin{eqnarray}
C_i(k^i;a) = a^{-2} G_i(ak^i).
\end{eqnarray}

2. The continuum limit of $C_i$ exists and is given by
\begin{eqnarray}
\lim_{a \to 0}C_i(k^i;a) = {(k^i)}^2.
\end{eqnarray}

3. There are two positive constants, $a_0$ and $A$, such that
\begin{eqnarray}
\left| C_i(k^i;a) \right| \ge A {(\hat k^i)}^2,
\end{eqnarray}
for all $a\le a_0$ and all $k^i$, where
$$ \hat k_\mu^i = {2 \over a}\sin({ak_\mu^i \over 2}). $$

Note that the free naive propagator verifies (20) and (21) but not (22).

It is not worthwhile to specify the requirements $V(k,q;a)$ must satisfy
because they are not very restrictive and are satisfied in all (local)
lattice models we know of.

Suppose, in addition, that the degree of divergence of $I_F$ [12]
is less than zero.

Then, the Reisz power counting theorem, states that the continuum limit of
$I_F$ exists and is given by
\begin{eqnarray}
\lim_{a \to 0}I_F = \int d^4k^1 \dots d^4k^L
{\lim\limits_{a \to 0}V(k,q;a) \over
\prod\limits_i\lim\limits_{a \to 0}C_i(k^i;a)},
\end{eqnarray}
where the integral of the r.h.s. is absolutely convergent.

The important modification incorporated by Reisz to the BPHZL
[13] approach consists in replacing Taylor operators by
Taylor polynomials in \lq\lq lattice momenta\rq\rq
\quad $( \sin(qa) / a)$.

Then he shows that the substractions can be written as counterterm
contributions
to the lattice action [14]. The method works for fields carrying
spin and internal symmetries like colour, etc.

The conclusion is that when the vertices and the propagators of a model in
the lattice verify the conditions of the Reisz theorem, then the contiuum
limit of any Green function calculated perturbatively with the model in the
lattice coincides, {\it after renormalization,}
with the Green function of the
model in the continuum (because of the interchange of limits in (23) ).

Next we will see that our lattice vertices and propagators satisfy, in an
effective sense, the conditions of the Reisz theorem.

Before doing that, it is enlightening to perform an explicit one--loop
computation of the fermion propagator(see appendix of ref. [11]). The result is
that the one--loop self--energy vanishes when the external momentum coincide
with one of the replicas. Consequently the position of replicas in momentum
space are not affected by the quantum corrections. This is an important point
if we want that our form factor, $F(\theta)$, will work beyond the tree level.
In fact we can see the proof of the decoupling, based on the Reisz theorem, as
a powerfull way of proving this to all orders in perturbation theory.

The reason why the theorem does not work in the case of the naive fermions is
that the denominator, $D_F(k,a) = (1 / a^2)\sum_\mu\sin^2(ak_\mu)$,
of the naive propagator is, for $k_\mu$ near $\pi / a$, of order one
(doubling problem) while $A{\hat k}^2$ can be arbitrarilly large for
$a \to 0$ independently of the value of $A$.

On the contrary, for our regularization this theorem works.

The reason is that, for any Feynman diagram, an internal propagator is always
accompanied by two $F(\theta)$ factors coming from the vertices. Then, in
the power counting theorem, the naive denominator is replaced by an
effective denominator, $\tilde D = {D_F(k,a) / F^2(ak)}$, which now
can be bounded by $A{\hat k}^2$ for all $k$
\footnote {The zeros of $F^2(ak)$ kill the zeros in $\sin(ak_\mu)$ for
$k_\mu = {\pi / a}$ and then both terms in (22) are of the same order in
$1 / a $. Then it is enough to take $A$ sufficiently small to satisfy the
bound.}.
Note that each vertex differs from the naive vertex in two $F(\theta)$
factors, one for each fermion field entering the vertex.

When applied to our case, the result we have obtained (i.e. the continuum limit
of any Green function coincides, after renormalization,
with the Green function of the model in the continuum limit with only one
fermion) means that the replica fermions do not give any contribution
to Green functions with finite external momenta.

Also, one can check performing an explicit one loop computation of the fermion
propagator, $S(p)$, that the behaviour of $S(p)$ around a point $\bar p$
($\bar p$ denotes the position of the replica poles) is the corresponding one
to the lattice naive free fermion propagator. In fact, defining
$ k = p - \bar p $, for $k = O(a^0)$ one finds
\begin{eqnarray}
S_{1-loop}(p) = {} -
{i\sum_\mu \gamma_\mu k_\mu \cos(\bar p_\mu a) \over \sum_\mu k_\mu^2 }
+ O(a),
\end{eqnarray}
which is the result one would expect if the decoupling has been achieved.

One last comment about the symmetries of the quantum theory defined by our
action (15). As our regularization preserves the global $U(1) \times U(1)$
symmetry, this symmetry is realized by the quantum theory.
Of course in this realization the massless decoupled replica fermions play
an essential r\^ole. Note that, under the $U(1) \times U(1)$ transformation,
\lq\lq the replica fermions\rq\rq
\footnote{ It is easy to check that the generalization of (7) which picks out
each replica is,
$$ \Psi^{(i)}(\theta) = \prod_\mu \cos({\theta_\mu + \theta_\mu^{(i)} \over 2})
\Psi(\theta),\qquad i = 1,\dots,2^d.$$
For $\theta_\mu^{(i)} = 0$ we obtain (7). Also
$\Psi_x = \sum\limits_i\Psi_x^{(i)}$.}
transform in the same way as the \lq\lq interacting $\Psi^{(1)}$ fermion\rq\rq.
Currents $\bar\Psi\gamma_\mu\Psi$ and $\bar\Psi\gamma_\mu\gamma_5\Psi$
are conserved as a consequence of cancellation of contributions of the
interacting and decoupled replica fermions. But the physical relevant currents,
made up from the $\Psi^{(1)}$ component, should not be simultaneously
conserved.

Let us finish up with a comment about other regularizations. One can ask
himself how the assumptions of the Reisz power counting theorem are satisfied
in other regularizarion approaches. Of course the Wilson propagator of the
Rome regularizarion satisfies those assumptions. The situation is different for
the model based on the Wilson--Yukawa term (Smit--Swift model
[1,3] for instance). In fact, in the paramagnetic phase with
weak Yukawa couplings (PMW) of a \cym, $\langle\Phi\rangle = 0$,
the propagator is the naive one and the vertex, $V$, does not have the
adequate behaviour to cause any definition of an effective propagator to
satisfy (22). Therefore, in this phase the decoupling is not assured.
Actually we know that one finds $2^d$ massless fermions in the PMW phase.
In the ferromagnetic phase with weak Yukawa couplings (FMW) one
recovers, because $v =: \langle\Phi\rangle \not= 0$, essentially the Wilson
propagator BUT, as we approach the phase transition line, $v$ approachs
zero, then the Reisz theorem is not satisfied
and fermion doubling occurs [1,15,16].

Of course, the nondecoupling effects of the doublers at one loop level found
by Dugan and Randall in other regularizarions [17], are not
present in our case as our decoupling method is not based on massive doubler
fermions.

\section{THE PHASE DIAGRAM FOR THE $U(1) \times U(1)$ CHIRAL YUKAWA MODEL}

For the moment, we have found, in a mean field computation, the phase structure
of the model defined by (15) for the case in which $\Phi\in U(1)$ and for
small $y$. We hope to complete this phase diagram, soon [18].
For a review of this topic see reference [19]. For a recent
comparison of the phase structure of different models see reference
[20].

Before showing our results, we must say something about the symmetries of the
phase diagram, because they are not exactly the same as in other
regularizations
and have relevant consequences.

We have some of the usual symmetries. For instance, for $y = 0$ the action is
invariant under $k \to -k$,\quad $\Phi_x \to \epsilon_x\Phi_x$,\quad
$\epsilon_x = {(-1)}^{x_1+x_2+x_3+x_4}$. For $y \not= 0$ it is invariant under
$$\Phi_x \to -\Phi_x \qquad y \to -y,$$ so we take $y\ge 0$ with no loss of
generality.

{\it But we have lost the symmetry $k \to -k\quad y \to -iy$, usual in other
models} [16,19].

In fact, in the implementation of this symmetry the term $I_B(\Phi)$ of the
action,
$$
I_B(\Phi) = {k \over 2}\sum_{x,\mu}(\Phi_x^+\Phi_{x+\hat\mu}
+ \Phi_{x+\hat\mu}^+\Phi_x),
$$
forces one to balance the change in $k$ with the change $\Phi_x \to
\epsilon_x\Phi_x$ (do not forget this). This implies a change in $I_Y$
which should be balanced, too. Obviously this change can not be balanced
with only a change in $y$, we need some change in $\Psi^{(1)}$ also.
This change in $\Psi^{(1)}$ must be induced by a change in $\Psi$ which
{\it should not} produce, in turn, any effect on $I_F(\Psi)$.
In other regularizations this is achieved by
\begin{eqnarray}
    \Psi_x \to e^{i\epsilon_x\alpha}    \Psi_x,\qquad
\bar\Psi_x \to e^{i\epsilon_x\alpha}\bar\Psi_x,
\end{eqnarray}
because, then,
$$\bar\Psi_x\Psi_{x + \hat\mu} \to
e^{i\alpha(\epsilon_x+\epsilon_{x+\hat\mu})}\bar\Psi_x\Psi_{x + \hat\mu} =
\bar\Psi_x\Psi_{x + \hat\mu}, \forall\alpha.$$
Usually, in other regularizations, there is a value of $\alpha(\pi/4)$
and a change in
$y\ (\to -iy)$ which keeps $I_Y$ unchanged. But, in our case, in
\begin{eqnarray}
\bar\Psi_x^{(1)}\Phi_x\Psi_x^{(1)} =
{1 \over 2^8}&(&\bar\Psi_x + \bar\Psi_{x+\hat 1}+ \bar\Psi_{x+\hat 2} + \dots)
\Phi_x \nonumber \\
             &(&\Psi_x + \Psi_{x+\hat 1}+ \Psi_{x+\hat 2} + \dots),
\nonumber
\end{eqnarray}
there are terms, such as $\bar\Psi_x\Phi_x\Psi_{x + \hat 1}$, which stay
unaffected by (25), but which change when $\Phi_x \to \epsilon_x\Phi_x$.
{\it Therefore, in our case, we do not have the symmetry
$k \to -k,\ y \to -iy$.}

The consequence of this fact is that \lq\lq the phase transition line
PM--AFM is not determined by the phase transition line FM--PM\rq\rq ,
as usually happens [16]. In our model, a simple mean field
computation yields that both lines meet in a point. We will return to
this point later on. Then, it is also possible that the mean field approach
yields a ferrimagnetic phase to the right of this point, making it a
quadruple point [18] (an interesting possibility already found in
numerical simulations [16,19]).

Also concerning the limiting case $y \to \infty$, we have a peculiarity
similar to that in the Chiral Yukawa model with hypercubic coupling
[21]. In fact, a rescaling of the fermion field
$ \Psi \to (1/\sqrt y)\Psi$, suppresses, as always, the fermion
kinetic term but, in our case, the term $I_Y$ also couples fermions in
different points (because of the spliting in $\Psi^{(1)}$) so that the
fermion can now propagate and it does not decouples in the limit
$y \to \infty$. So, the phase diagram is not,
{\it in this limit,} that of a pure $\Phi^4$ model.

The next step is to make a mean field analysis of the phase diagram.

The more significant caracteristic of our mean field computation are:

1) We have taken the saddle point approach to the mean field calculation
[22].

2) To handle the four fermion interaction which occurs when developing, for
small $y$, the term $e^{-I_Y}$, we must, before doing the integration over
the fermi fields, decouple the composite field $\bar\Psi(x)\Psi(x)$
by using the identity [23]
$$ \exp\{{1\over 2}{(\bar\Psi_x\Psi_x)}^2 V_x^{-1}\} =
{[\det(V_x)]}^{1\over 2} $$
$${} \times \int {d\lambda(x)\over\sqrt{2\pi}}
\exp\{-{1\over 2}{\lambda(x)}^2 V_x + \lambda(x)\bar\Psi_x\Psi_x\}.$$

The phase diagram for small $y$ is shown in figure 2.

\begin{figure}[htbp]
\vspace {88mm}
\caption{
Phase diagram for the $ U(1) \times U(1) $ chiral Yukawa model for small
$y$ and $\Phi \in U(1)$. The large and intermediate values of $\bar y$ will
be considered in ref. [18] ($\bar y = y/\protect\sqrt 2$).}
\end{figure}

The FM--PM phase transition is second order and the transition line is,
$$
2k_cd = 1 - {\bar y}^2 2^{d \over 2}
\int {d^d\theta \over {(2\pi)}^d}
{F^4(\theta) \over \sum_\lambda\sin^2\theta_\lambda}.
$$
For $d = 4$,
\begin{eqnarray}
\int {d^4\theta \over {(2\pi)}^4}
{F^4(\theta) \over \sum_\lambda\sin^2\theta_\lambda} = 1.61 \times {10}^{-2}.
\end{eqnarray}

The PM--AFM phase transition is also second order and the transition line is,
$$
2k_cd = -1 - {\bar y}^2 2^{d \over 2}
\int {d^d\theta \over {(2\pi)}^d}
{F^2(\theta) F^2(\theta + \pi) \over \sum_\lambda\sin^2\theta_\lambda}.
$$
For $d = 4$,
\begin{eqnarray}
\int {d^4\theta \over {(2\pi)}^4}
{F^2(\theta) F^2(\theta + \pi) \over \sum_\lambda\sin^2\theta_\lambda}
 = 8.4 \times {10}^{-5},
\end{eqnarray}
so that $k_c$ changes very little with $y$.

The reason why these two lines meet in a point is that in (26) $F^4(\theta)$
kills the most important contribution of the replica while in (27)
$F^2(\theta) F^2(\theta + \pi)$ kills the replica AND the \lq\lq normal\rq\rq\
fermions, which implies a very small value for the integral.

In the naive case, $$ F^4(\theta) = F^2(\theta) F^2(\theta + \pi) = 1, $$
we recover the $k \to -k \quad y \to -iy$ symmetry and both lines are parallel.

Note that, in the regularized theory, the value of $\bar y$ in which the two
lines meet decreases as the width of the form factor, $F(\theta)$, decreases.
Of course, in a hypothetical scaling limit the physical quantities should be
the same for any form factor which kills the replicas.

Now, we are studying the intermediate and the large $y$ values considering, in
our mean field analysis, the general situation of a ferrimagnetic phase.
To solve the saddle point equations in this general situation is rather
involved [18].

\section{OUTLOOK}

Almost everything is still to be done:

i) Estimation of the upper bounds on fermion masses (and comparison with the
symmetry breaking scale). This question may be relevant to the real world.
{}From LEP data and comparison with radiative corrections in the SM one
concludes that if $m_t < 250\ G\eV$ (to make sure that the one loop
calculations
are dominant) then $100 G\eV < m_t < 150\ G\eV$. However those perturbative
analysis are not valid for very heavy top. In fact, for $y_t^2/(4\pi^2)
> 1$ the contribution to $\Delta\rho\quad (\rho = 1 + \Delta\rho;\quad \rho$
measures the relative strengh of charged-current and neutral-current effective
couplings) beyond the perturbative regime is not known. Then, the present
precision weak interaction measurements could be compatible with very big
values of $m_t$.

At this point a comment on the decoupling theorem of Appelquist and Carazzone
[24] is pertaining.

This theorem does not hold in the PERTURBATIVE regime of gauge theories with
spontaneous symmetry breaking (SSB). This happens because of the presence of
large Yukawa couplings which grow with heavy fermion masses. In a way, this
is nice because it allows us to see the effect of heavy fermions at low
energies. But, on the other hand, it is surprising that a fermion of,
for instance, infinite mass may have significance in real life. Perhaps
the decoupling theorem of Appelquist and Carazzone also holds in the presence
of SSB but, in order to notice it, one must go beyond the perturbative regime.

Well, we hope to compute the contributions to $\Delta\rho$ which are
responsible
for the perturbative bounds on $m_t$
{\it beyond the perturbative regime,} to see
if these contributions grow up or go down with $y_t$.

ii) Of course, one must study the phase diagram for other CYM, as
$ SU(2) \times SU(2)$, for instance.

iii) The bounds on the Higgs mass should also be reexamined in the presence of
heavy fermions.

iv) Also the connection of CYM and four fermion interaction models should be
investigated.

v) At the level of a CGT all we know is that the one loop perturbative analysis
from the Rome approach can be directly applied to our case: everything else
remains to be done. Nevertheless we hope that the global chiral symmetry
present with this action will make things a little easier.

\begin{acknowledge}
We wish to thank the organizers for giving us the opportunity to participate in
this very stimulating workshop. We thank to D. Espriu and B. Grinstein for
very interesting discussions. This work was partially supported by the
CICYT (proyecto AEN 90-0030).
\end{acknowledge}

\end{document}